\documentclass{iau} 
\usepackage{graphicx}

\title[Bright-red stars in the Leo~A galaxy] 
{Bright-red stars in the dwarf irregular galaxy Leo~A}

\author[A. Le\v{s}\v{c}inskait\.{e}, R. Stonkut\.{e} \& V. Vansevi\v{c}ius]   
{Alina Le\v{s}\v{c}inskait\.{e}$^1$, Rima Stonkut\.{e}$^{1,2}$, \and  Vladas Vansevi\v{c}ius$^{1,2}$}

\affiliation{$^1$Center for Physical Sciences and Technology, \\ Saul\.{e}tekis av. 3, 10257 Vilnius, Lithuania \\email: {\tt alina.lescinskaite@ftmc.lt} \\[\affilskip]
$^2$Astronomical Observatory of Vilnius University, \\ M. K. \v{C}iurlionis st. 29, 03100 Vilnius, Lithuania \\email: {\tt vladas.vansevicius@ff.vu.lt}
}

\pubyear{2018}
\volume{344}  
\jname{Dwarf Galaxies: From the Deep Universe to the Present}
\editors{A.C. Editor, B.D. Editor \& C.E. Editor, eds.}
\begin{document}

\newcommand{\farcm}{\mbox{\ensuremath{.\mkern-4mu^\prime}}}

\maketitle

\begin{abstract}
We analysed a population of bright-red (BR) stars in the dwarf irregular galaxy Leo~A by using multicolour photometry data obtained with the Subaru/Suprime-Cam ($B$, $V$, $R$, $I$, $H\alpha$) and HST/ACS ($F475W$ \& $F814W$) instruments. In order to separate the Milky Way (MW) and Leo~A populations of red stars, we developed a photometric method, which enabled us to study the spatial distribution of BR stars within the Leo~A galaxy. 

We found a significant difference in the scale-length (S-L) of radial distributions of the ``young" and ``old" red giant branch (RGB) stars -- $0\farcm82 \pm 0\farcm04$ and $1\farcm53 \pm 0\farcm03$, respectively. Also, we determined the S-L of BR stars of $0\farcm85 \pm 0\farcm05$, which closely matches that of the ``young" RGB stars. Additionally, we found a sequence of peculiar RGB stars and 8 dust-enshrouded stars in the Leo~A galaxy. 

\keywords{galaxies: dwarf, galaxies: individual (Leo~A), galaxies: stellar content}
\end{abstract}

\firstsection 
\section{Introduction}

According to dwarf galaxy evolution scenarios, star formation is predominantly confined to the inner regions, where denser gas tends to reside. This leads to prominent age gradients forming over time in these galaxies, as younger objects concentrate in the centre and the older ones spread out to larger radii, see, e.g., \cite[McQuinn et al. (2017)]{McQuinn17}. While tidal interactions and gas accretion can clearly affect the distributions of stellar populations in dwarf galaxies, on the other hand, stellar feedback and star migration may also reduce real age gradients (\cite[El-Badry et al. 2016]{ElBadry16}). This suggests that a detailed study of the radial extent of different stellar populations within dwarf irregular galaxies might provide some insight into their dynamical and morphological evolution.

\section{Foreground/Background Decontamination}

\begin{figure}
\begin{center}
 \includegraphics[width=4in]{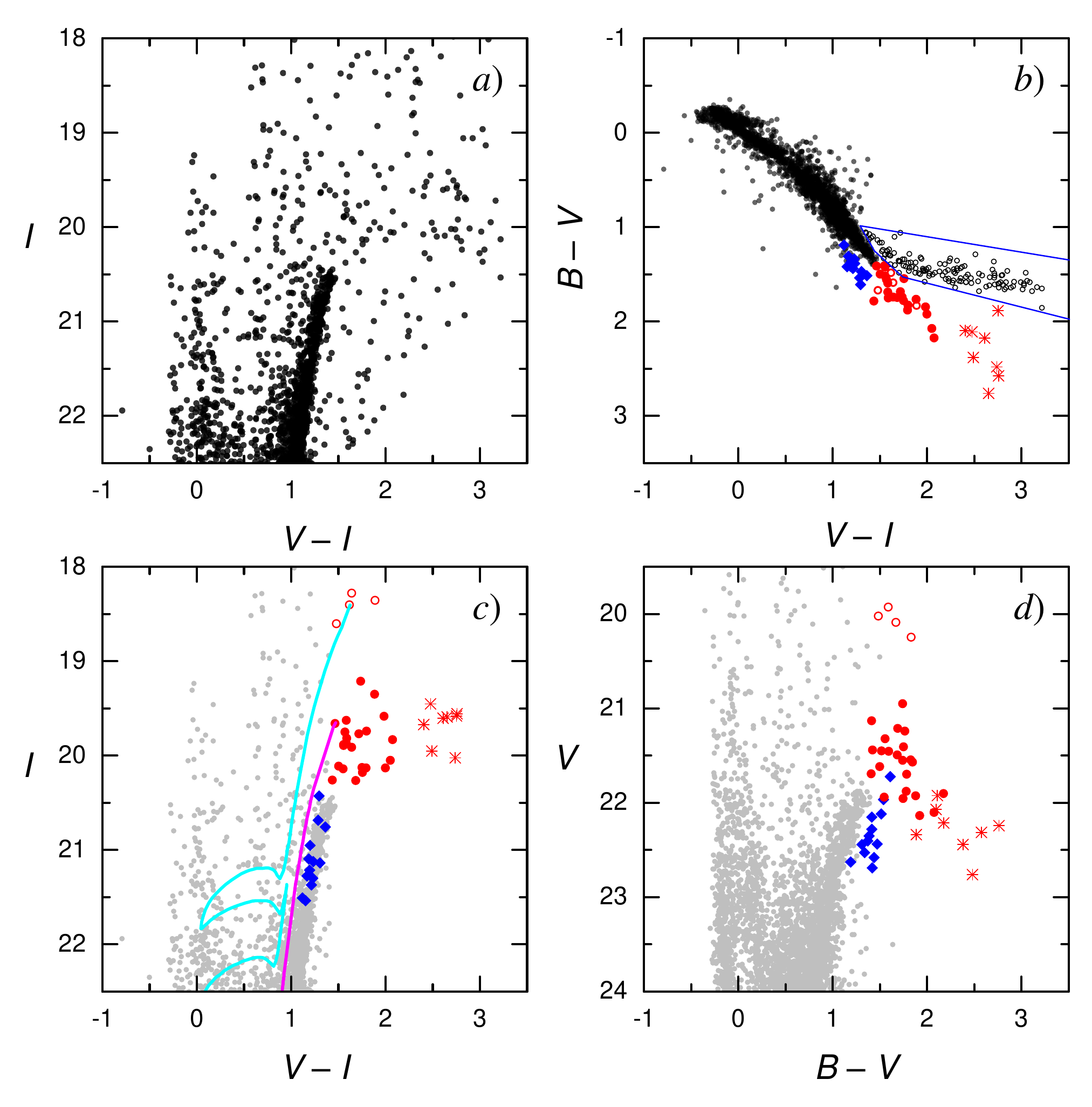} 
 \caption{a) Initial CMD: $I$, $V - I$. b) Initial two-colour diagram; black open circles (enclosed by blue lines) mark foreground MW dwarfs of K-M spectral types. c) Decontaminated CMD: $I$, $V - I$; overlaid are PARSEC isochrones (\cite[Marigo et al. 2017]{Marigo17}) of ages 0.25 (cyan) \& 1.0 (magenta) Gyr ($Z=0.0007$). d) Decontaminated CMD: $V$, $B - V$. b-d) Red symbols mark BR stars: younger stars -- open circles; older -- filled circles; older dusty -- asterisks; peculiar RGB stars -- blue diamonds. }
   \label{fig1}
\end{center}
\end{figure}

The catalogue of star photometry (\cite[Stonkut\.{e} et al. 2014]{Stonkute14}) based on Subaru Suprime-Cam imaging data of the Leo~A galaxy contains a large number of foreground and background contaminants. In order to clean out the catalogue, all bright objects, $V < 24$ (N~=~5240), were visually inspected using multicolour Subaru and HST/ACS images. As a result, $\sim$30\% of objects were identified as being non-stellar (i.e., background galaxies, unresolved star blends, or star images distorted by bright neighboring objects) and removed from the catalogue. Then the catalogue was decontaminated of K-M spectral type foreground MW dwarfs by excluding a distinct sequence of stars (corresponding well to the isochrones of solar metallicity) in a two-colour diagram, $B-V$ vs. $V-I$ (Fig.\,\ref{fig1}b). The result of decontamination is evident by comparing colour-magnitude diagrams (CMDs) Fig.\,\ref{fig1}a\&c.

\section{Red Stars in the Leo~A galaxy}

\begin{figure}
\begin{center}
 \includegraphics[width=4.5in]{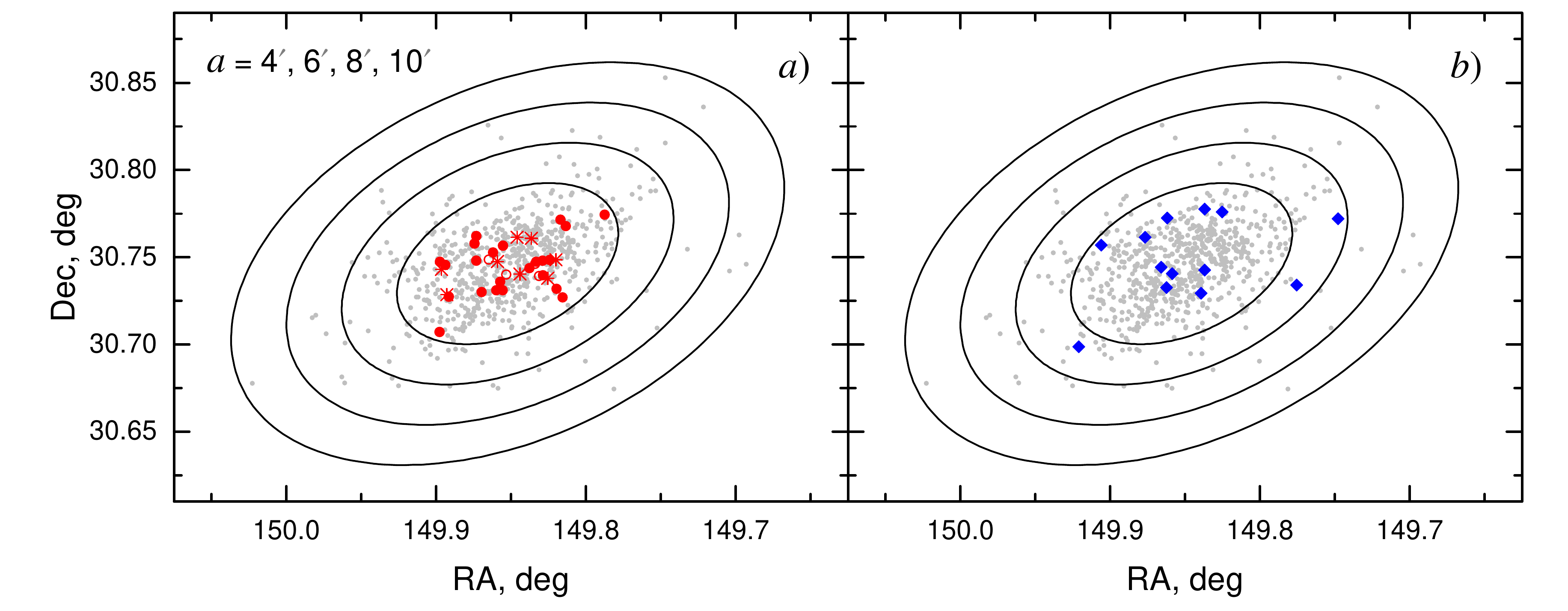} 
 \caption{Spatial distributions of BR (a) and peculiar RGB (b) stars. Coloured symbols are the same as in Fig.\,\ref{fig1}b-d, grey points mark the RGB stars (Fig.\,\ref{fig3}a). Depicted ellipses are as follows: $a$ -- semi-major axis; ellipticity, $b/a = 0.6$; position angle of the major axis, P.A.~$=115^{\circ}$.}
   \label{fig2}
\end{center}
\end{figure}

\begin{figure}
\begin{center}
 \includegraphics[width=5.4in]{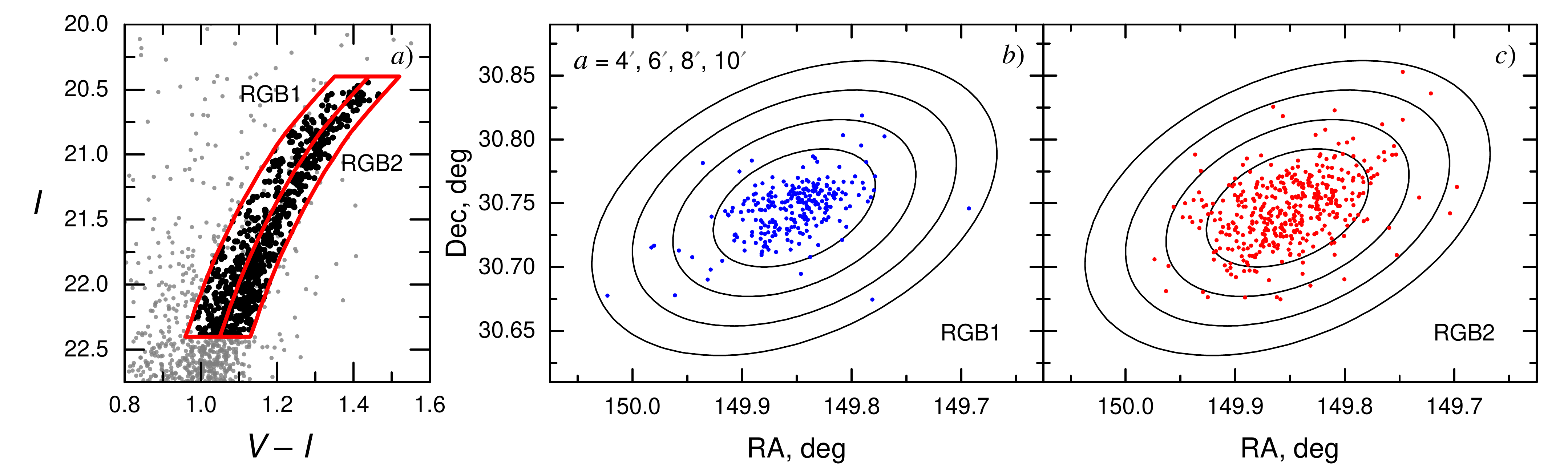} 
 \caption{a) CMD of the RGB stars (black points, enclosed by red lines); the RGB stars are divided into two subsets by the lines closely matching an isochrone of 5~Gyr, $Z=0.0005$: RGB1 (left) and RGB2 (right). b\&c) Spatial distributions of the RGB stars as selected in panel (a). Depicted ellipses are the same as in Fig.\,\ref{fig2}.}
   \label{fig3}
\end{center}
\end{figure}

BR and RGB stars in the Leo~A galaxy were selected based on their positions in CMD, $I$ vs. $V-I$ (Fig.\,\ref{fig1}c). The stars brighter and redder than the tip of RGB (TRGB, $I = 20.4$) were classified as BR. The selected sample of BR stars (Fig.\,\ref{fig2}a) consists of 4 objects ($I < 19$) representing a younger generation of stars, $\lesssim$300~Myr, and 31 objects ($19.0 < I < 20.4$) that are older than $\sim$1~Gyr (Fig.\,\ref{fig1}c). We further analyse the older population only.

The selected sample of RGB stars extends only two magnitudes below TRGB (Fig.\,\ref{fig3}a) in order to avoid contamination by the red-clump stars. Contamination by non-RGB objects was reduced by applying the following photometric criteria: $Q_{BVI} > (16.57 - I) / 16.67$, $Q_{BVI} < (21.57 - I) / 16.67$, and $-0.05 < H\alpha-R < 0.05$, where $Q_{BVI}=(B-V)-(V-I)$.

\section{Results and Discussion}

\begin{figure}
\begin{center}
 \includegraphics[width=3.2in]{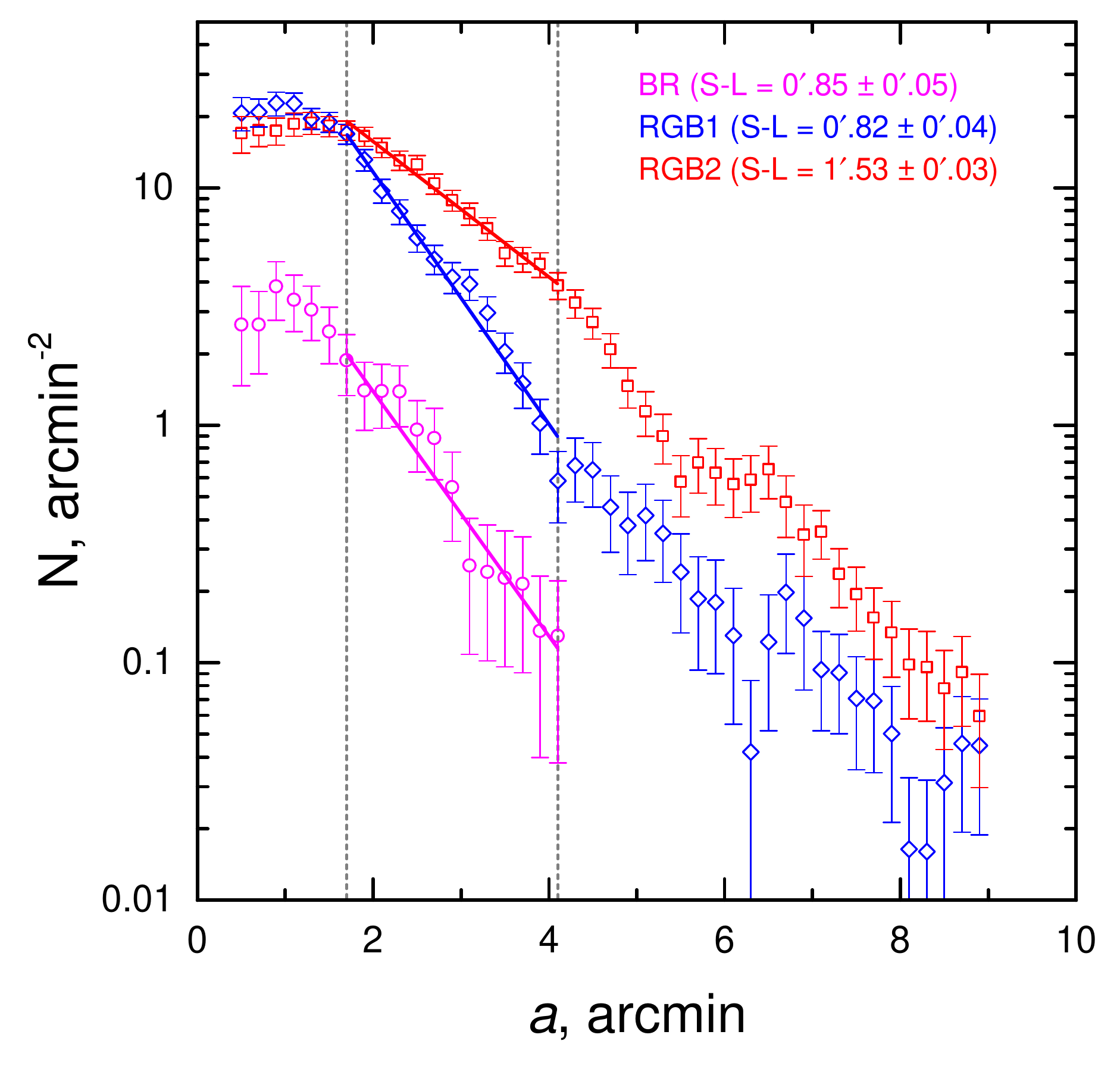} 
 \caption{Surface number density profiles of RGB (RGB1 -- blue diamonds, RGB2 -- red squares) and BR (magenta circles) stars measured in elliptical annuli. Linear fits are shown for each profile and the corresponding S-L are indicated in the legend. The fits are confined to a region indicated by the vertical dashed lines. The inner radius limit is chosen to avoid a crowded central part of Leo~A (completeness at $I = 22.5$ mag is $\sim$80\%). The outer limit coincides with the extent of the BR stars.}
   \label{fig4}
\end{center}
\end{figure}

By dividing the RGB sequence into the blue and red parts (RGB1 and RGB2, respectively; Fig.\,\ref{fig3}a) we found distinct radial distributions (Figs.\,\ref{fig3}b \&\,\ref{fig3}c) with corresponding S-L of $0\farcm82 \pm 0\farcm04$ and $1\farcm53 \pm 0\farcm03$ (Fig.\,\ref{fig4}). The distribution of BR stars (S-L = $0\farcm85 \pm 0\farcm05$) matches well that of RGB1 population, suggesting that these stars might belong to the same population.

Assuming that delayed star formation (with a peak $\sim$5~Gyr ago; \cite[Skillman et al. 2014]{Skillman14}) resulted in insignificant changes in metallicity during the early star forming history of the Leo~A galaxy (\cite[Kirby et al. 2017]{Kirby17}), the age might be the main factor affecting the width of the RGB sequence, thus making the majority of the RGB1 stars representative of the younger population, while most of the RGB2 stars are likely to be older. 

Distinct structural components with notable differences in age and spatial extent imply an outside-in star formation or efficient star migration to the outskirts of the Leo~A galaxy in early stages of its evolution. Together with previously identified stellar populations at $5' < a < 8'$ (\cite[Vansevi\v{c}ius et al. 2004]{Vansevicius04}), this might indicate that even isolated small size and low mass galaxies, like Leo~A, have a non-trivial star formation history and structure evolution scenario driven solely by internal processes of stellar evolution.

We also identified a group of 13 peculiar objects fully overlapping with the RGB stars in the $I$ vs. $V-I$ diagram (Fig.\,\ref{fig1}c), yet forming a separate sequence in diagrams with the $B$ passband (Fig.\,\ref{fig1}b\&d). However, this sequence is not apparent in the HST/ACS data. Photometry data of these stars are not contaminated by neighboring objects or defects. The peculiar objects are projected on the Leo~A galaxy (Fig.\,\ref{fig2}b), and are expected to be genuine members. Their unusual properties might be a consequence of differences in chemical composition. This is implied by two of the peculiar stars that have spectroscopic observations (\cite[Kirby et al. 2017]{Kirby17}), as both fall in the outskirts of metallicity distribution, with [Fe/H] values of $-0.57\pm0.11$ and $-0.88\pm0.11$, as compared to the average of $-1.67^{+0.09}_{-0.08}$. However, spectroscopy of a larger sample is needed in order to confirm that.

Cross-matching the optical photometry catalogue with the near-infrared (NIR) data obtained by \cite[Jones et al. (2018)]{Jones18} with the WIYN High-resolution Infrared Camera (WHIRC), we find counterparts for 20 of our bright-red stars (11 stars fall outside the WHIRC's field). Five of these stars have dust production rates (DPR) $>10^{-11}$ M\textsubscript{\(\odot\)} yr$^{-1}$ and are thus considered to be dust-producing stars. Four of these five dust-enshrouded stars are found within a somewhat distinct group of 8 BR stars with $V-I > 2.3$ mag and $V$ magnitudes dimmer than the TRGB (Fig.\,\ref{fig1}b-d), implying that the other four stars in this group are also likely to have dusty envelopes.
\\

$Acknowledgements$. This research was funded by a grant No. LAT-09/2016 from the Research Council of Lithuania.

\end{document}